%% file: Auger_EnergySpectrum.tex
\documentclass[final,3p,times,twocolumn]{elsarticle}

\usepackage{graphicx}
\usepackage{amssymb}
\usepackage{amsthm}
\usepackage{units}
\usepackage{color}
\usepackage{booktabs}

\journal{Physics Letters}

\newcommand{\comment}[1]{}

\begin{document}

\begin{frontmatter}

\title{Measurement of the energy spectrum of cosmic rays above
$\unit[10^{18}]{eV}$ using the Pierre Auger Observatory}

\author[auger]{The Pierre Auger Collaboration}
\address[auger]{Observatorio Pierre Auger, Av. San Martin Norte 304, 5613 Malarg\"ue, Argentina}

\address{}

\begin{abstract}

  We report a measurement of the flux of cosmic rays with
  unprecedented precision and statistics using the Pierre Auger
  Observatory.  Based on fluorescence observations in coincidence with
  at least one surface detector we derive a spectrum for energies
  above $\unit[10^{18}]{eV}$.  We also update the previously published
  energy spectrum obtained with the surface detector array.  The two
  spectra are combined addressing the systematic uncertainties and, in
  particular, the influence of the energy resolution on the spectral
  shape. The spectrum can be described by a broken power law
  $E^{-\gamma}$ with index $\gamma=3.3$ below the ankle which is measured at
  $\log_{10}(E_\mathrm{ankle}/\mathrm{eV}) = 18.6$. Above the ankle
  the spectrum is described by a power law with index $2.6$ followed
  by a flux suppression, above about $\log_{10}(E/\mathrm{eV}) = 19.5$, 
  detected with high statistical significance.

\end{abstract}

\end{frontmatter}

\onecolumn
\input author_list_latex.tex
\twocolumn

\section{Introduction}

The flux of ultra-high energy cosmic rays exhibits two important
features. At energies above $\unit[4 \times 10^{19}]{eV}$ a
suppression of the flux with respect to a power law extrapolation is
found \cite{Abbasi:2007sv,Abraham:2008ru}, which is compatible with
the predicted Greisen-Zatsepin-Kuz'min (GZK) effect
\cite{Greisen:1966jv,Zatsepin66}, but could also be related to the
maximum energy that can be reached at the sources. A break in the
power law, called the ankle, is observed at an energy of about
$\unit[3 \times 10^{18}]{eV}$
\cite{Linsley:1963m1,Lawrence:1991cc,Nagano:1992jz,Bird:1993yi}.  This
break in the energy spectrum has traditionally been attributed to the
transition from the galactic component of the cosmic ray flux to a
flux dominated by extragalactic sources
\cite{Hillas:2005cs,Wibig:2004ye}. In recent years it became clear
that a similar feature in the cosmic ray spectrum could also result
from the propagation of protons from extragalactic sources, placing
the transition from galactic to extragalactic cosmic rays at a much
lower energy \cite{Berezinsky:2005cq,Aloisio:2007rc}. 
In this model the ankle is produced by the modification of the source
spectrum of primary protons. This is caused by $e^\pm$ pair production
of protons with the photons of the cosmic microwave background,
leading to a well-defined prediction of the shape of the flux in the
ankle region.
 
Accurate measurement of the
cosmic ray flux in the ankle region is expected to help determine the
energy range of the transition between galactic and extragalactic
cosmic rays and to constrain model scenarios.

Two complementary techniques are used at the Pierre Auger Observatory
to detect extensive air showers initiated by ultra-high energy cosmic
rays (UHECR): a {\it surface detector array (SD)} and a {\it
  fluorescence detector (FD)}. The SD of the southern observatory in
Argentina consists of an array of 1600 water Cherenkov detectors
covering an area of about $\unit[3000]{km^2}$ on a triangular grid
with $\unit[1.5]{km}$ spacing. Electrons, photons and muons in air
showers are sampled at ground level with a on-time of almost
$\unit[100]{\%}$. In addition the atmosphere above the surface
detector is observed during clear, dark nights by 24 optical
telescopes grouped in 4 buildings. These detectors are used to observe
the longitudinal development of extensive air showers by detecting the
fluorescence light emitted by excited nitrogen molecules and the
Cherenkov light induced by the shower particles. Details of the design
and status of the Observatory are given
elsewhere~\cite{Abraham:2004dt, Allekotte:2007sf, AugerFDPaper}.

The energy spectrum of ultra-high energy cosmic rays at energies
greater than $\unit[2.5 \times 10^{18}]{eV}$ has been derived using
data from the surface detector array of the Pierre Auger
Observatory~\cite{Abraham:2008ru}. This measurement provided evidence
for the suppression of the flux above $\unit[4 \times 10^{19}]{eV}$
and is updated here. In this work we extend the previous measurements
to lower energies by analysing air showers measured with the
fluorescence detector that also triggered at least one of the stations
of the surface detector array.  Despite the limited event statistics
due to the fluorescence detector on-time of about $\unit[13]{\%}$, the
lower energy threshold and the good energy resolution of these {\it
  hybrid} events allow us to measure the flux of cosmic rays in the
region of the ankle.

The energy spectrum of hybrid events is determined from data taken
between November 2005 and May 2008, during which the Auger Observatory
was still under construction. Using selection criteria that are set
out below, the exposure accumulated during this period was computed
and the flux of cosmic rays above $\unit[10^{18}]{eV}$ determined. 
The spectrum obtained with the surface detector array, updated using data
until the end of December 2008, is combined with the hybrid one to 
obtain a spectrum measurement over a wide energy range with the 
highest statistics available.

\section{Hybrid energy spectrum}

The hybrid approach to shower observation is based on the shower
detection with the FD in coincidence with at least one SD station. The
latter condition, though insufficient to establish an independent SD
trigger~\cite{Abraham:2008ru, SDacceptance}, enables the shower
geometry and consequently the energy of the primary particle to be
determined accurately. The reconstruction accuracy of hybrid events is
much better than what can be achieved using SD or FD data
independently \cite{Fick:2003qp+Sommers:1995dm}. For example, the
energy resolution of these hybrid measurements is better than $6\%$
above $\unit[10^{18}]{eV}$ compared with about $15\%$ for the surface
detector data.\\ \indent Event reconstruction proceeds in two
steps. First the shower geometry is found by combining information
from the shower image and timing measured with the FD with the trigger
time of the surface detector station that has the largest signal
\cite{Mostafa:2006id}. In the second step the profile of energy deposition 
of the shower is reconstructed \cite{Unger:2008uq} and
shower parameters such as depth of shower maximum and primary particle
energy are calculated together with their uncertainties.

\subsection{Event selection and reconstruction}

 To ensure good energy reconstruction only events that satisfy the
following quality criteria are accepted:
\begin{itemize}
\item Showers must have a reconstructed zenith angle smaller than $ \unit[60]{^\circ}$.
\item In the plane perpendicular to the shower axis, the reconstructed
 shower core must be within $\unit[1500]{m}$ of the station used for
 the geometrical reconstruction.
\item
The contribution of Cherenkov light to the overall signal of the FD
must be less than $\unit[50]{\%}$.
\item 
The Gaisser-Hillas fit~\cite{ Unger:2008uq, Gaisser77a} of the reconstructed
 longitudinal profile must be successful with $\chi^{2}$/Ndof $<$
 2.5.
\item
The maximum of the shower development, $X_{\rm max}$, must be
observed in the field of view of the telescopes.
\item
The uncertainty in the reconstructed energy, which includes light flux
 and geometrical uncertainties, must be $\sigma(E)/E<\unit[20]{\%}$.
\item Only periods during which no clouds were detected above the Observatory
are used.
\end{itemize}

To avoid a possible bias in event selection due to the differences
between shower profiles initiated by primaries of different mass, only
showers with geometries that would allow the observation of all
primaries in the range from proton to iron are retained in the data
sample. The corresponding fiducial volume in shower-telescope distance
and zenith angle range is defined as a function of the reconstructed
energy and has been verified with data~\cite{Abraham:2006ar}. About
$1700$ events fulfil the selection criteria for quality and for 
fiducial volume.\\ \indent

A detailed simulation of the detector response has shown that every FD
trigger above $E= \unit[10^{18}]{eV}$ passing all the described
selection criteria is accompanied by a SD trigger of at least one
station, independent of the mass and direction of the incoming primary
particle~\cite{Perrone:2007he}.

\subsection{Exposure calculation}\label{sec:exposure}
During the time period discussed here the southern Auger Observatory
was in its construction phase with the number of available SD stations
increasing from around 630 to a nearly fully completed
instrument with 1600 detectors. 
Over the same period the FD was enlarged from 12 to 24
telescopes. In addition to these large scale changes, smaller but 
important changes occur on much shorter timescales due,
for example, to hardware failures. The data-taking of the fluorescence
detector is furthermore influenced by weather effects such as storms or
rainfall. These and other factors that affect the efficiency of the data-taking
must be taken into account in the determination of the aperture.

The total exposure is the integral over the instantaneous
aperture and can be written as
\begin{equation}
 {\mathcal E}(E) =  \int_{T} \int_{\Omega} \int_{S_{\mathrm{gen}}} \varepsilon(E, t,
  \theta, \phi, x,
  y)~\cos\theta~\textrm{d}S~\textrm{d}\Omega~\textrm{d}t,
\end{equation}
where $\textrm{d}\Omega = \sin\theta\textrm{d}\theta\textrm{d}\phi$
and $\Omega$ are respectively the differential and total solid angles,
$\theta$ and $\phi$ are the zenith and azimuth angles and
$\textrm{d}S=\textrm{d}x \times \textrm{d}y$ is the horizontal surface
element. The final selection efficiency $\varepsilon$ includes the
efficiencies of the various steps of the analysis, namely the trigger,
reconstruction and selection efficiencies and also the evolution of
the detector during the time period $T$. It has been derived from
Monte Carlo simulations that scan an area $S_{\mathrm{gen}}$ large
enough to enclose the full detector array.

\begin{figure}[t!]
  \includegraphics[width=\linewidth]{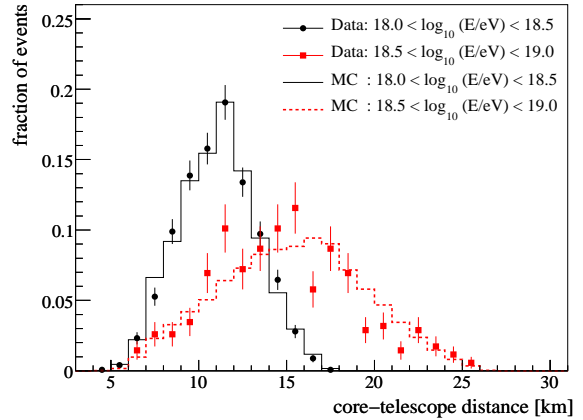}
  \caption{Distribution of events observed with the fluorescence
  detector as a function of the distance of the shower core from the
  telescopes for data and Monte Carlo simulation.\label{fig:CoreEyeDistance}}
\end{figure}

The changing configuration of the SD array is taken into account for
the determination of the hybrid on-time. In addition, within time
intervals of $\unit[10]{min}$, the status of all detector components
of the Pierre Auger Observatory down to the level of single PMTs of
the fluorescence detector is determined. Moreover all known
inefficiencies such as DAQ read-out deadtimes are considered.

The longitudinal profile of the deposition of energy simulated with
the QGSJet-II~\cite{Ostapchenko:2005nj,Ostapchenko:2006vr} and
Sibyll 2.1~\cite{Fletcher:1994bd,Engel:1999db} hadronic interaction
models and the CONEX~\cite{Bergmann:2006yz} air shower simulation
program are the basis for an extensive set of Monte Carlo
simulations. The exact data taking conditions are reproduced by means
of a detailed detector simulation within the Auger analysis
framework~\cite{Argiro:2007qg}. All atmospheric measurements,
e.g. scattering and absorption lengths, as well as monitoring
information such as the noise caused by night sky background light and
PMT trigger thresholds are taken into account. 

The reconstruction of the simulated showers is then performed in
exactly the same way as for the data and good agreement between data and
Monte Carlo simulations is obtained. As an example, we show in
Fig.~\ref{fig:CoreEyeDistance} the distribution of events observed
with the fluorescence detector as a function of the distance of the
shower core from the telescopes.

\begin{figure}[t!]
  \includegraphics[width=\linewidth]{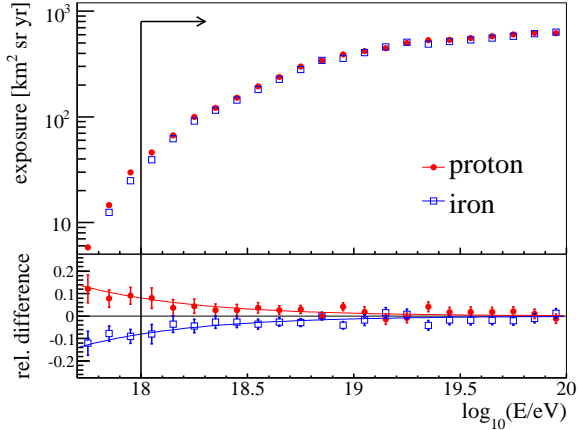}
  \caption{The hybrid exposure for different primary particles
  together with the difference to the mixed composition used for the
  flux measurement.
 \label{fig:Exposure}}
\end{figure}

 Fig.~\ref{fig:Exposure} shows the hybrid exposure of events
fulfilling all of the quality and fiducial volume cuts that have been 
applied, for proton and iron primaries. As can be seen, the cuts adopted 
lead to only a small dependence of the exposure on the mass composition 
which can be assumed to be dominated by hadrons~\cite{AugerNeutrinoLimit2009, AugerPhotonLimit}. 
The systematic uncertainty arising from our lack of knowledge of the mass 
composition is about $8\%$ at $\unit[10^{18}]{eV}$ and decreases to less 
than $1\%$ above $\unit[10^{19}]{eV}$. 
We assume a mixed composition of $\unit[50]{\%}$ proton and $\unit[50]{\%}$ 
iron nuclei for the flux determination and include the remaining composition 
dependence in the systematic uncertainty. 
The dependence of the exposure on the assumed model of hadronic interactions 
was found to be less than 2\% over all the energy range.

The full MC simulation chain has been cross-checked with air shower
observations and the analysis of laser shots that are fired from the
Central Laser Facility~\cite{Arqueros:2006zz} and detected with the
fluorescence detector. Following this analysis the exposure has been 
reduced by $8\%$ to account for lost events and an upper limit to 
the remaining systematic uncertainty of $\unit[5]{\%}$
was derived~\cite{ICRC09_Salamida}. By combination with the
uncertainty related to mass composition the total systematic
uncertainty of the hybrid exposure is estimated as $\unit[10]{\%}$
($\unit[6]{\%}$) at $\unit[10^{18}]{eV}$ ($>\unit[10^{19}]{eV}$).

\subsection{Energy spectrum from hybrid data}
The flux of cosmic rays $J$ as a function of energy is given by

\begin{equation}
  J(E) = \frac{\mathrm{d}^4N_{\mathrm{inc}}}{\mathrm{d}E\,
    \mathrm{d}A\, \mathrm{d}\Omega\, \mathrm{d}t } \\ \cong
    \frac{\Delta N_{\mathrm{sel}}(E)}{\Delta E}\frac{1}{{\mathcal
    E}(E)}\;,
\end{equation}
where $N_{\mathrm{inc}}$ is the number of cosmic rays with energy $E$
incident on a surface element $\mathrm{d}A$, within a solid angle
$\mathrm{d}\Omega$ and time $\mathrm{d}t$. $\Delta
N_{\mathrm{sel}}(E)$ is the number of detected events passing the
quality cuts in the energy bin centered around $E$ and having width
$\Delta E$. ${\mathcal E}(E)$ is the energy-dependent exposure defined
above.

\begin{figure}[!t]
  \includegraphics[width=\linewidth]{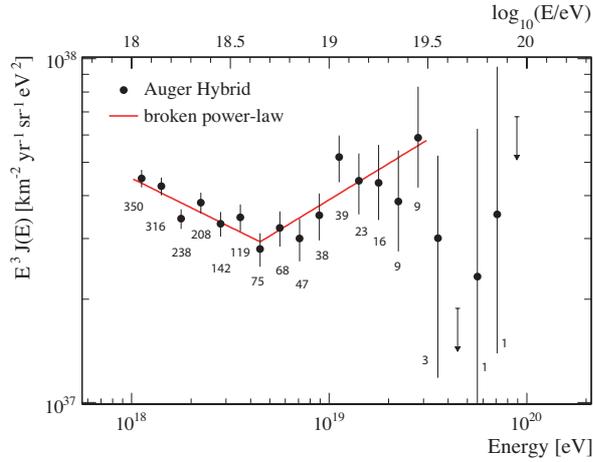}
  \caption{The energy spectrum of ultra-high energy cosmic rays
    determined from hybrid measurements of the Pierre Auger
    Observatory. The number of events is given for each of the energy
    bins next to the corresponding data point. Only statistical
    uncertainties are shown. The upper limits correspond to the $68\%$
    CL. A fit with a broken power law is used to determine the
    position of the ankle.}\label{fig:HDSpectrum}
\end{figure}

The measured flux as function of energy is shown in
Fig.~\ref{fig:HDSpectrum}. A break in the power law of the derived
energy spectrum is clearly visible. The position of this feature, known as the
ankle, has been determined by fitting two power laws $J = k
E^{-\gamma}$ with a free break between them in the energy interval
from $\unit[10^{18}]{eV}$ to $\unit[10^{19.5}]{eV}$. The upper end of
this interval was defined by the flux suppression observed in the
spectrum derived using surface detector
data~\cite{Abraham:2008ru}. The ankle is found at
$\log_{10}(E_{\mathrm{ankle}}/\mathrm{eV})=18.65 \pm 0.09{\rm (stat)}
{}^{+ 0.10}_{-0.11} {\rm (sys)}$
and the two power law indices have been determined as 
$\gamma_{\mathrm{1}}= 3.28 \pm 0.07 {\rm (stat)} {}^{+0.11}_{-0.10}
{\rm (sys)}$
and
$\gamma_{\mathrm{2}}=2.65 \pm 0.14 {\rm (stat)} {}^{+0.16}_{-0.14}
{\rm (sys)}$,
($\chi^2/\mathrm{ndof} = 10.2/11$),
where the systematic uncertainty is due to the residual effect of 
the unknown mass composition.

The energy estimation of fluorescence measurements relies on the
knowledge of the fluorescence yield. Here we adopt the same absolute
calibration~\cite{Nagano:2003zn} and the wavelength and pressure
dependence~\cite{Ave:2007xh} as in Ref.~\cite{Abraham:2008ru}.  This
is currently one of the dominant sources of systematic uncertainty
($\unit[14]{\%}$). The fraction of the energy of the primary particle
that is carried by muons and neutrinos and does not contribute to the
fluorescence signal has been calculated based on air shower
simulations and goes from about $\unit[14]{\%}$ at $\unit[10^{18}]{eV}$ 
to about $\unit[10]{\%}$ at $\unit[10^{19}]{eV}$~\cite{Barbosa:2004}. 
The systematic uncertainty depending on the choice of models and mass
composition is about $\unit[8]{\%}$~\cite{Pierog:2007x2}. Further
systematic uncertainties in the absolute energy scale are related to
the absolute detector calibration ($\unit[9.5]{\%}$) and its
wavelength dependence
($\unit[3]{\%}$)~\cite{Knapik:2007yd}. Uncertainties of the lateral
width of the shower image and other reconstruction uncertainties
amount to about $\unit[10]{\%}$ systematic uncertainty in the energy
determination. Atmospheric conditions play a crucial role for air
shower observations with fluorescence detectors. An extensive program 
of atmospheric monitoring is conducted at the Pierre Auger
Observatory allowing the determination of the relevant parameters and
the associated
uncertainties~\cite{Arqueros:2006zz,BenZvi:2006xb,Keilhauer:2005ja,BenZvi:2007it}.
The total systematic uncertainty in the energy determination is
estimated as $\unit[22]{\%}$~\cite{ICRC09_DiGiulio}. Indirect methods
of determining the energy scale, which do not involve the fluorescence
detector calibration, seem to indicate an energy normalization that is
higher than the one used here by an amount comparable to the
uncertainty given above~\cite{ICRC09_Castellina}.

\section{Update of Surface Detector Spectrum}
Here we update the published energy spectrum based on surface detector
data~\cite{Abraham:2008ru} using data until the end of December 2008. 
The exposure is now $\unit[12,790]{km^2~sr~yr}$. The event selection 
requires that the detector station with the highest
signal be surrounded by operational stations and that the
reconstructed zenith angle be smaller than
$60^\circ$~\cite{SDacceptance}. More than $35,000$ events fulfill
these criteria.

The energy estimator of the surface detector is corrected for shower
attenuation effects using a constant-intensity method. The
calibration of this energy estimator with fluorescence measurements
has been updated using the increased data set of high-quality hybrid
events~\cite{ICRC09_DiGiulio}.

Because of the energy resolution of the surface detector data (about
$20\%$ at the lowest energies, improving to about $10\%$ at the
highest energies), bin-to-bin migrations influence the reconstruction
of the flux and spectral shape. To correct for these effects, a
forward-folding approach is applied. MC simulations are used to
determine the energy resolution of the surface detector and a
bin-to-bin migration matrix is derived. The matrix is then used to
find a flux parameterisation that matches the measured data after
forward-folding. The ratio of this parameterisation to the folded flux
gives a correction factor that is applied to the data. The correction
to the flux is mildly energy dependent and is less than $20\%$ over
the full energy range. Details will be discussed in a forthcoming
publication.

\begin{figure}[t!]
  \includegraphics[width=\linewidth]{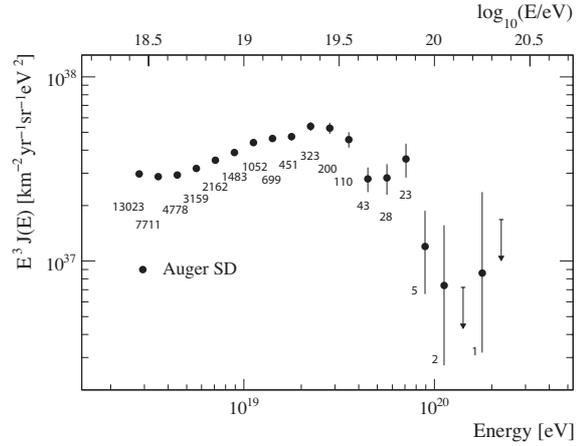}
  \caption{Energy spectrum, corrected for energy resolution, derived
    from surface detector data calibrated with fluorescence
    measurements. The number of events is given for each of the energy
    bins next to the corresponding data point. Only statistical
    uncertainties are shown. The upper limits correspond to $68\%$
    CL.}\label{fig:SDSpectrum}
\end{figure}

The energy spectrum, after correction for the energy resolution, is
shown in Fig.~\ref{fig:SDSpectrum} together with the event numbers of
the underlying raw distribution. Combining the systematic
uncertainties of the exposure ($3\%$) and of the forward folding
assumptions ($5\%$), the systematic uncertainty of the derived flux
is $6\%$.

\section{The Combined Auger spectrum}
 The energy spectrum derived from hybrid data is combined with the one
obtained from surface detector data using a maximum likelihood
method. Since the surface detector energy estimator is calibrated with
hybrid events, the two spectra have the same systematic uncertainty in
the energy scale. On the other hand, the normalisation uncertainties
are independent. They are taken as $\unit[6]{\%}$ for the SD and
$\unit[10]{\%}$ ($\unit[6]{\%}$) for the hybrid flux at
$\unit[10^{18}]{eV}$ ($>\unit[10^{19}]{eV}$). These normalisation
uncertainties are used as additional constraints in the
combination. This combination procedure is used to derive the scale 
parameters, $k$, for the fluxes that are to be applied to the individual spectra. 
These are $k_{\mathrm{SD}}=1.01$ and $k_{\mathrm{FD}}=0.99$ for the surface
detector data and hybrid data respectively, showing that agreement
between the measurements is at the 1\% level.

\begin{figure*}[thb!]
\begin{center}  
\includegraphics[width=0.75\linewidth]{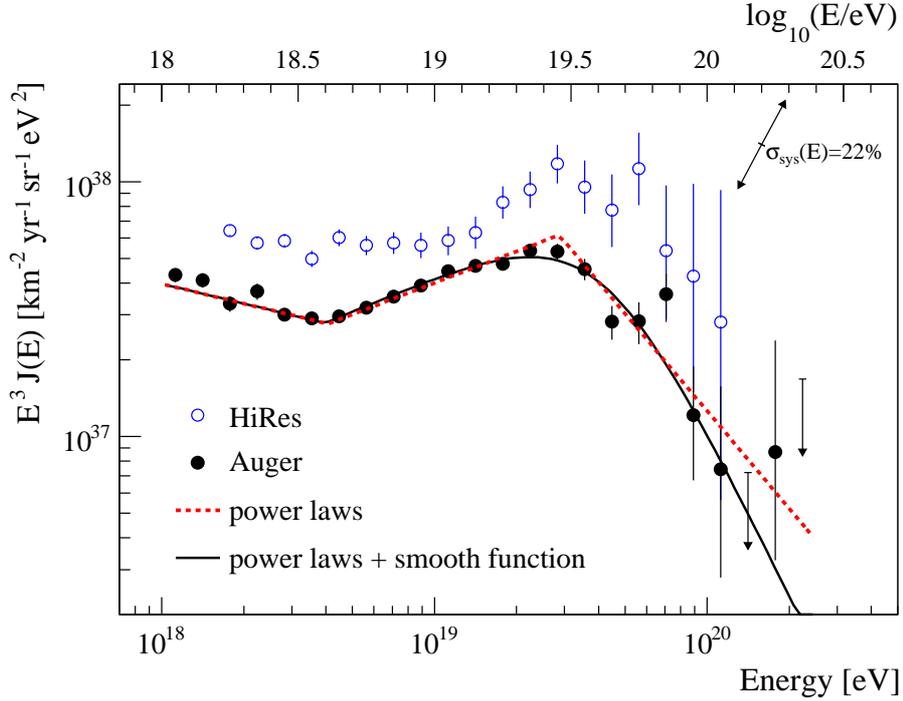}
\vspace*{-4mm}
\end{center}
\caption{The combined energy spectrum is fitted with two functions
  (see text) and compared to data from the HiRes
  instrument~\cite{HiResStereoSpectrum}. The systematic uncertainty of
  the flux scaled by $E^3$ due to the uncertainty of the energy scale of 22\% is
  indicated by arrows. A table with the Auger flux values can be found
  at~\cite{SpectrumDataWeb}.
  \label{fig:SpectrumComparison} }
\end{figure*}

The combined energy spectrum scaled with $E^{3}$ is shown in
Fig.~\ref{fig:SpectrumComparison} in comparison with the spectrum
obtained with stereo measurements of the HiRes
instrument~\cite{HiResStereoSpectrum}. An energy shift within the
current systematic uncertainties of the energy scale applied to one or
both experiments could account for most of the difference between the
spectra. The ankle feature seems to be somewhat more sharply defined
in the Auger data. This is possibly due to a systematic energy offset
between the experiments. However, for a complete comparison, care must
also be taken to account for energy resolution and possible changes in
aperture with energy.

\begin{table}[!t]
\caption[]{Fitted parameters and their statistical uncertainties
  characterising the combined energy spectrum.}\label{tab:fit}
\begin{center}
\begin{tabular}{lrr} \toprule
parameter & power laws & power laws \\
 & & + smooth function \\ \midrule
$\gamma_1 (E < E_\mathrm{ankle})$ & $3.26 \pm 0.04$ & $3.26 \pm 0.04$
\\ 
$\log_{10}(E_\mathrm{ankle}/\mathrm{eV})$ & $18.61 \pm 0.01$ & $18.60 \pm
0.01$ \\ 
$\gamma_2 (E > E_\mathrm{ankle})$ & $2.59 \pm 0.02$ &  $2.55 \pm 0.04$
\\ 
$\log_{10}(E_\mathrm{break}/\mathrm{eV})$ & $19.46 \pm 0.03$ & \\ 
$\gamma_3 (E > E_\mathrm{break}) $ & $4.3 \pm 0.2$ &  \\ 
$\log_{10}(E_\mathrm{\nicefrac{1}{2}}/\mathrm{eV})$ &  & $19.61 \pm 0.03$\\ 
$\log_{10}(W_\mathrm{c}/\mathrm{eV})$ &  & $0.16 \pm 0.03$ \\
$\chi^2/\mathrm{ndof}$ & $38.5 / 16 $ & $29.1 / 16$ \\ 
 \bottomrule
\end{tabular}%
\end{center}
\end{table}

The characteristic features of the combined spectrum are quantified in
two ways. For the first method, shown as a dotted red line in
Fig.~\ref{fig:SpectrumComparison}, we have used three power laws with
free breaks between them. A continuation of the power law above the
ankle to highest energies can be rejected with more than $20$
$\sigma$. For the second characterisation we have adopted two power laws
in the ankle region and a smoothly changing function at higher
energies which is given by
\begin{equation}
J(E;E>E_{\mathrm{ankle}}) \propto \frac{E^{-\gamma_2}}{ 1 +
\exp\left(\frac{\log_{10}{E}-\log_{10}{E_{\nicefrac{1}{2}}}}{\log_{10}{W_c}}\right) },
\end{equation}
where $E_{\nicefrac{1}{2}}$ is the energy at which the flux has fallen
to one half of the value of the power-law extrapolation and $W_c$
parametrizes the width of the transition region.  It is shown as a black
solid line in Fig.~\ref{fig:SpectrumComparison}. The derived
parameters (quoting only statistical uncertainties) are given in
Tab.~\ref{tab:fit}.

At high energies the combined spectrum is statistically dominated by
the surface detector data. The agreement between the index of the
power law above the ankle, $\gamma_2$, measured with the combined
spectrum ($2.59\pm 0.02$) and with hybrid data ($2.65\pm 0.14$), also
demonstrates the good agreement between the two measurements.

\section{Summary}

We have measured the cosmic ray flux with the Pierre Auger Observatory
by applying two different techniques. The fluxes obtained with hybrid
events and from the surface detector array are in good agreement in
the overlapping energy range. A combined spectrum has been derived
with high statistics covering the energy range from
$\unit[10^{18}]{eV}$ to above $\unit[10^{20}]{eV}$. The dominant
systematic uncertainty of the spectrum stems from that of the overall
energy scale, which is estimated to be $22\%$.

The position of the ankle at $\log_{10}(E_\mathrm{ankle}/\mathrm{eV})
= 18.61~\pm~0.01$ has been determined by fitting the flux with a
broken power law $E^{-\gamma}$. An index of $\gamma = \unit[3.26 \pm
0.04]{}$ is found below the ankle. Above the ankle the spectrum
follows a power law with index $\unit[2.55 \pm 0.04]{}$. In comparison
to the power law extrapolation, the spectrum is suppressed by a factor
two at $\log_{10}(E_{\nicefrac{1}{2}}/\mathrm{eV})=\unit[19.61 \pm
0.03]{}$. The significance of the suppression is larger than
20\,$\sigma$.  The suppression is similar to what is expected from the
GZK effect for protons or nuclei as heavy as iron, but could in part
also be related to a change of the shape of the average injection
spectrum at the sources.

\vspace{7mm}
\noindent {\bf Acknowledgements.}
\input acknowledgements.tex

\bibliographystyle{unsrt-mod-notitle}

\end{document}

%% file: author_list_latex.tex
\par\noindent
{\bf The Pierre Auger Collaboration} \\
J.~Abraham$^{8}$, 
P.~Abreu$^{73}$, 
M.~Aglietta$^{55}$, 
E.J.~Ahn$^{89}$, 
D.~Allard$^{31}$, 
I.~Allekotte$^{1}$, 
J.~Allen$^{92}$, 
J.~Alvarez-Mu\~{n}iz$^{80}$, 
M.~Ambrosio$^{48}$, 
L.~Anchordoqui$^{106}$
% ...

\par\noindent
$^{1}$ Centro At\'{o}mico Bariloche and Instituto Balseiro (CNEA-
UNCuyo-CONICET), San Carlos de Bariloche, Argentina \\
% ...
$^{8}$ National Technological University, Faculty Mendoza 
(CONICET/CNEA), Mendoza, Argentina \\
% ...
$^{31}$ Laboratoire AstroParticule et Cosmologie (APC), 
Universit\'{e} Paris 7, CNRS-IN2P3, Paris, France \\
% ...
$^{48}$ Universit\`{a} di Napoli ``Federico II'' and Sezione INFN, 
Napoli, Italy \\
% ...
$^{55}$ Istituto di Fisica dello Spazio Interplanetario (INAF),
 Universit\`{a} di Torino and Sezione INFN, Torino, Italy \\
% ...
$^{73}$ LIP and Instituto Superior T\'{e}cnico, Lisboa, Portugal \\
% ...
$^{80}$ Universidad de Santiago de Compostela, Spain \\
% ...
$^{89}$ Fermilab, Batavia, IL, USA \\
% ...
$^{92}$ New York University, New York, NY, USA \\
% ...
$^{106}$ University of Wisconsin, Milwaukee, WI, USA \\
% ...
\par\noindent
(\ddag) Deceased\\
(a) at Konan University, Kobe, Japan\\
(b) On leave of absence at the Instituto Nacional de Astrofisica, Optica y Electronica\\
(c) at Caltech, Pasadena, USA\\

%% file: acknowledgements.tex
The successful installation and commissioning of the Pierre Auger Observatory
would not have been possible without the strong commitment and effort
from the technical and administrative staff in Malarg\"ue.

We are very grateful to the following agencies and organizations for financial support: 
Comisi\'on Nacional de Energ\'ia At\'omica, 
Fundaci\'on Antorchas,
Gobierno De La Provincia de Mendoza, 
Municipalidad de Malarg\"ue,
NDM Holdings and Valle Las Le\~nas, in gratitude for their continuing
cooperation over land access, Argentina; 
the Australian Research Council;
Conselho Nacional de Desenvolvimento Cient\'ifico e Tecnol\'ogico (CNPq),
Financiadora de Estudos e Projetos (FINEP),
Funda\c{c}\~ao de Amparo \`a Pesquisa do Estado de Rio de Janeiro (FAPERJ),
Funda\c{c}\~ao de Amparo \`a Pesquisa do Estado de S\~ao Paulo (FAPESP),
Minist\'erio de Ci\^{e}ncia e Tecnologia (MCT), Brazil;
AVCR AV0Z10100502 and AV0Z10100522,
GAAV KJB300100801 and KJB100100904,
MSMT-CR LA08016, LC527, 1M06002, and MSM0021620859, Czech Republic;
Centre de Calcul IN2P3/CNRS, 
Centre National de la Recherche Scientifique (CNRS),
Conseil R\'egional Ile-de-France,
D\'epartement  Physique Nucl\'eaire et Corpusculaire (PNC-IN2P3/CNRS),
D\'epartement Sciences de l'Univers (SDU-INSU/CNRS), France;
Bundesministerium f\"ur Bildung und Forschung (BMBF),
Deutsche Forschungsgemeinschaft (DFG),
Finanzministerium Baden-W\"urttemberg,
Helmholtz-Gemeinschaft Deutscher Forschungszentren (HGF),
Ministerium f\"ur Wissenschaft und Forschung, Nordrhein-Westfalen,
Ministerium f\"ur Wissenschaft, Forschung und Kunst, Baden-W\"urttemberg, Germany; 
Istituto Nazionale di Fisica Nucleare (INFN),
Ministero dell'Istruzione, dell'Universit\`a e della Ricerca (MIUR), Italy;
Consejo Nacional de Ciencia y Tecnolog\'ia (CONACYT), Mexico;
Ministerie van Onderwijs, Cultuur en Wetenschap,
Nederlandse Organisatie voor Wetenschappelijk Onderzoek (NWO),
Stichting voor Fundamenteel Onderzoek der Materie (FOM), Netherlands;
Ministry of Science and Higher Education,
Grant Nos. 1 P03 D 014 30, N202 090 31/0623, and PAP/218/2006, Poland;
Funda\c{c}\~ao para a Ci\^{e}ncia e a Tecnologia, Portugal;
Ministry for Higher Education, Science, and Technology,
Slovenian Research Agency, Slovenia;
Comunidad de Madrid, 
Consejer\'ia de Educaci\'on de la Comunidad de Castilla La Mancha, 
FEDER funds, 
Ministerio de Ciencia e Innovaci\'on,
Xunta de Galicia, Spain;
Science and Technology Facilities Council, United Kingdom;
Department of Energy, Contract Nos. DE-AC02-07CH11359, DE-FR02-04ER41300,
National Science Foundation, Grant No. 0450696,
The Grainger Foundation USA; 
ALFA-EC / HELEN,
European Union 6th Framework Program,
Grant No. MEIF-CT-2005-025057, 
European Union 7th Framework Program, Grant No. PIEF-GA-2008-220240,
and UNESCO.